\documentclass[journal]{IEEEtran}

\usepackage{xcolor}

\usepackage{amsmath,amssymb,cite,epsfig}
\usepackage{epsfig,amsmath,amssymb,epsf,cite,algorithm,algorithmic}
\usepackage{subfigure}
\usepackage{enumitem} 

\usepackage{hyperref}

\def\b0{{\pmb{0}}}

\begin{document}

\title{Key Focus Areas and Enabling Technologies for 6G}

\author{\IEEEauthorblockN{
Christopher G. Brinton,
Mung Chiang,
Kwang Taik Kim,
David J. Love,
Michael Beesley, \\
Morris Repeta,
John Roese,
Per Beming,
Erik Ekudden,
Clara Li,
Geng Wu, \\
Nishant Batra,
Amitava Ghosh,
Volker Ziegler,
Tingfang Ji,
Rajat Prakash,
John Smee
}
\thanks{This is an output of collaborations between Purdue, Cisco, Dell, Ericsson, Intel, Nokia, and Qualcomm. Portions were released online as a report ``6G Roadmap: A Global Taxonomy'' in Nov. 2023. The six companies are alphabetically listed, with authors from each institution in alphabetical order.}
\thanks{This work was supported in part by the National Science Foundation (NSF) under grants ITE-2326898, EEC-1941529, and CNS-2212565.}
}

\maketitle

\begin{abstract}
  We provide a taxonomy of a dozen enabling network architectures, protocols, and technologies that will define the evolution from 5G to 6G.
  These technologies span the network protocol stack, different target deployment environments, and various perceived levels of technical maturity.
  We outline four areas of societal focus that will be impacted by these technologies, and overview several research directions that hold the potential to address the problems in these important focus areas.
\end{abstract}
\vspace{-0.05in}

\vspace{-0.15in}
\section{Introduction}
\label{sect_intro}
Over the last decade, the availability of high-rate wireless communication has become a necessity, elevating the issue of future standards to international importance.
To fulfill the ever-growing number of connected devices and their demand for a diverse range of intelligent services, both technical and public policy innovations for the next generation are needed.

The discussion on 6G is well underway, and internationally distributed. Relevant groups include the Next G Alliance in the US, Hexa-X in the EU, Bharat 6G Alliance in India, China Academy of Information and Communications Technology (CAICT), XG Mobile Promotion Forum (XGMF) in Japan, and Telecommunications Technology Association (TTA) in Korea. While interest is high, there is still a lack of consensus on what 6G research and development should accomplish.

While existing 6G vision articles provide deep dives into specific technologies (e.g., \cite{li20226g,6g_leo}), we present a broader overview of 6G that covers its new frontiers, standardization activities, and societal implications of its emerging technologies. In particular, we shed light on technologies ranging from radio enhancements, such as new frequency bands and signal improvements to support higher data rates, to network deployment changes, which could facilitate changes to ways operators install and use base stations, to new computation and software solutions. Looking ahead to the 6G era, it becomes necessary to continue capturing the implications of these emerging technologies and analyze these new frontiers that promise to evolve 5G systems and disrupt their current trends.

\vspace{-0.15in}
\subsection{Societal Focus Areas}
\label{ssec:societal}
We identify four high-level focus areas that represent the challenges and solution needs for 6G development activities:

\textbf{(1) Scalability:}
Both the volume of data and number of devices are projected to continue increasing exponentially.
6G must accommodate an unknown future of demand, device and network heterogeneity, managing the dramatic increase in computation \textit{and} communication needed for artificial intelligence and machine learning (AI/ML) services.
Research must facilitate increased network spectral efficiencies and proper load balancing. We must figure out how to use a broad range of frequencies, including designing networks that ``play nice" with legacy users and efficiently use non-traditional bands. 

\textbf{(2) Sustainability:}
As devices and networks continue to operate in unprecedented ways, energy efficiency becomes even more essential.
The obvious requirement is lowering on-device and network-side power consumption. Waveform and medium access control innovations are needed to conserve battery life without hurting user experience, targeting zero energy consumed with zero traffic.
Research on joint design of distributed computing and communications protocols will improve energy efficiency for data-intensive applications. 6G will also enable numerous autonomy and power distribution applications, requiring sophisticated control frameworks \cite{NGA6Groadmap}.

\textbf{(3) Trustworthiness:}
The widespread availability of devices, the diversity in their production, and the massive increase in sensitive user data has given rise to security concerns not previously felt in commercial wireless. 6G must focus on a “built-in” approach to security, driving new approaches to network implementation and deployment \cite{5Gsecurity}. Distrust in hardware and software may be mitigated through open architectures, which minimize dependence on unverifiable and opaque compute and communication blocks. Techniques must be developed to harden 6G networks in a provable way.

\textbf{(4) Digital Inclusion:}
$\sim$40\% of the world's population remains unconnected, including significant portions of developed countries \cite{itu2021}.
Research will be needed to overcome economic and political challenges that have stifled expansion of reliable, high-rate wireless coverage to rural and third-world regions.
This could include research to enable growth in high altitude transmission point/base station deployments. Network designs that enable new backhaul techniques and increased spectral efficiencies are also needed for long-range coverage.

\vspace{-0.15in}
\subsection{Outline of 6G Technologies}
\begin{figure}
\epsfig{file=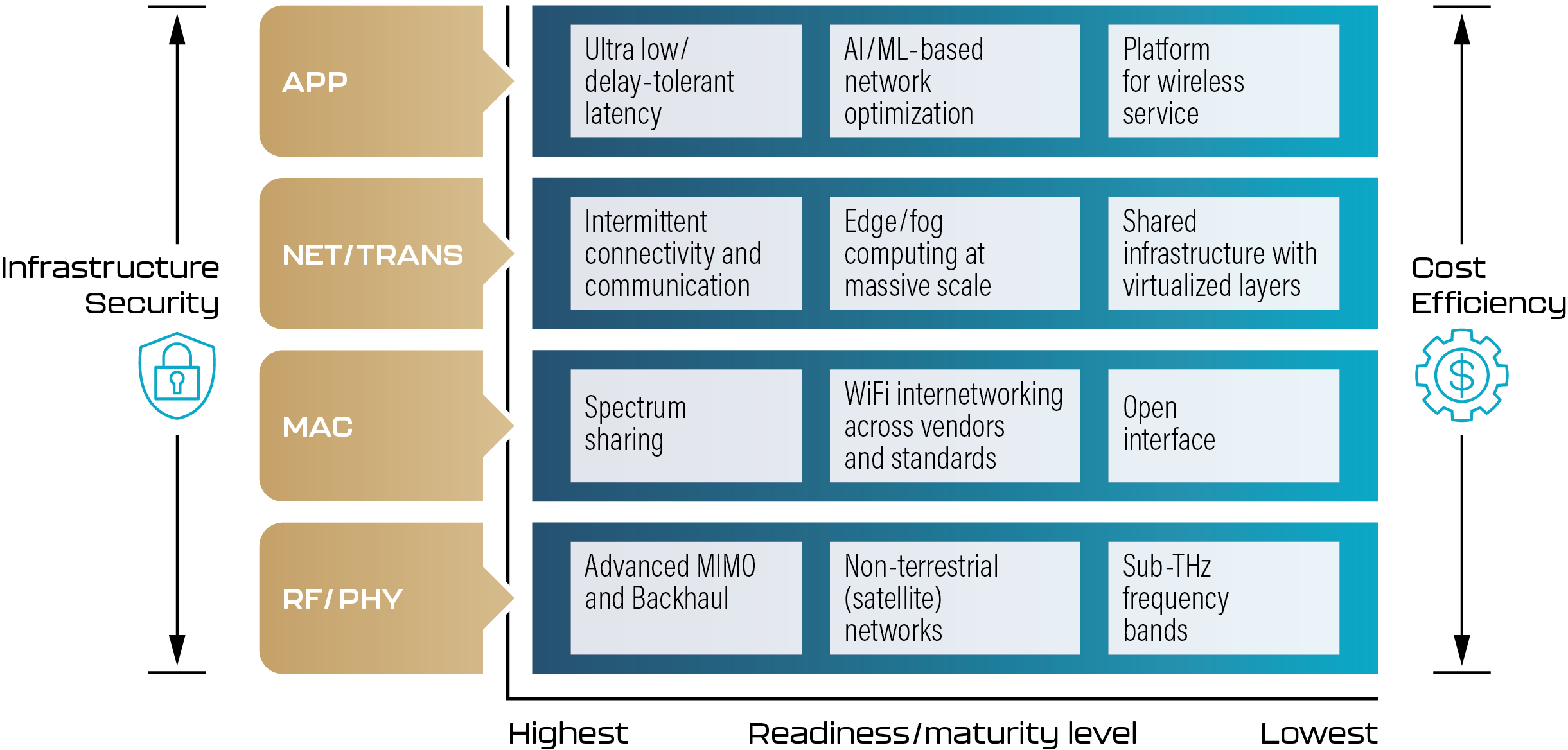,width=3.5in} \\[0.1in]
\centerline{\epsfig{file=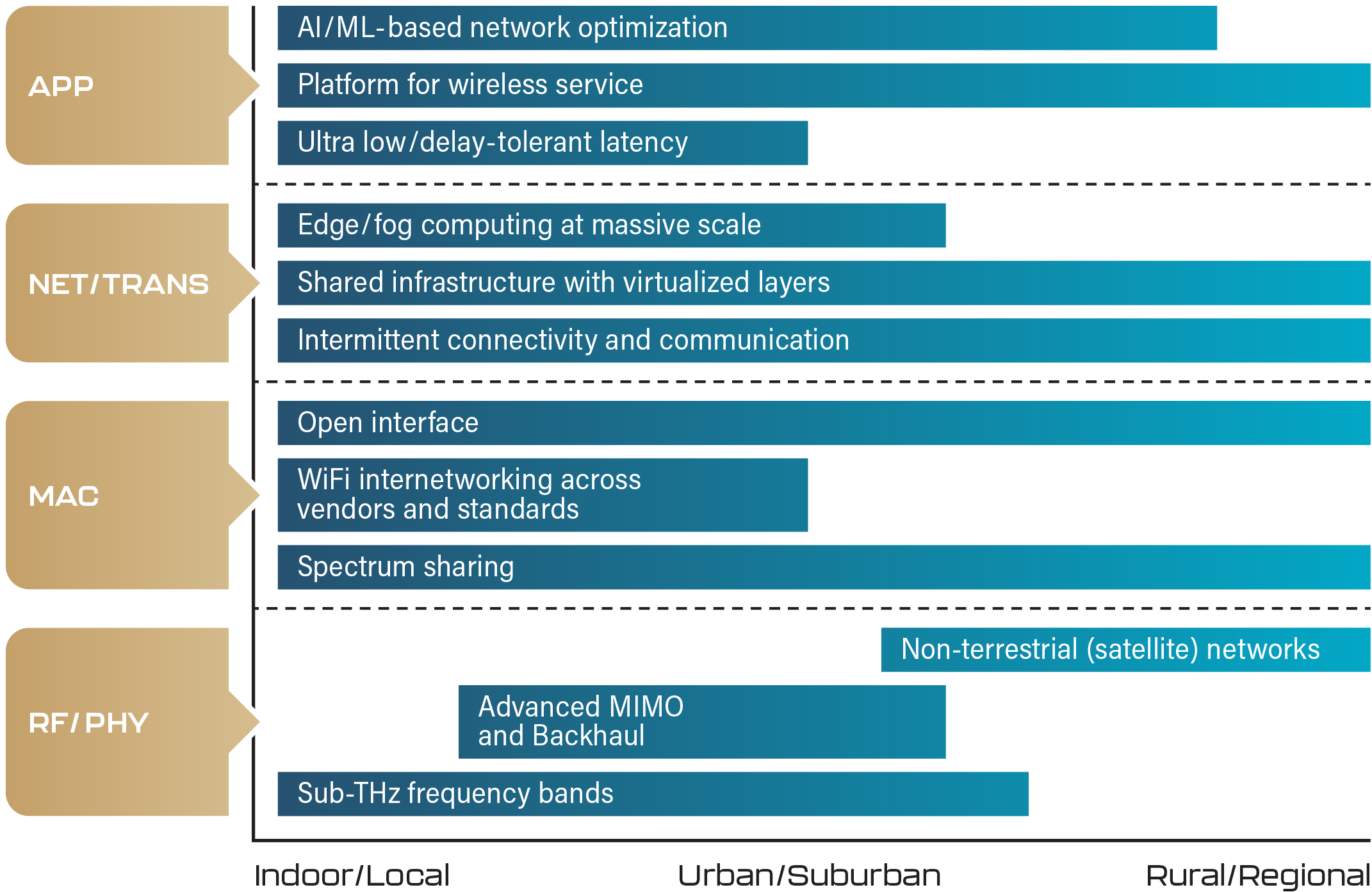,width=3.25in}}
\vspace{-0.25in}
\caption{6G technical focus areas organized by primary protocol layer, perceived technical readiness, and target deployment environment.}
\vspace{-0.2in}
\label{fig:6gtaxonomy}
\end{figure}

Fig. \ref{fig:6gtaxonomy} summarizes our partitioning of 12 6G-enabling technologies by primary protocol layer, perceived maturity level, and primary deployment environment. Sec. II-V group and discuss these technologies in technical thrusts, including their societal implications and whether they are expected in 6G's near-term (early 2030s), mid, or far-term (late 2030s) releases.
\vspace{-0.1in}
\section{New Radio Access and Connectivity}

\subsection{More MIMO}
Each new release of 3GPP has generally been accompanied by an increase in the number of downlink antenna ports \textcolor{black}{to support {scalability} demands.}
This allows for advanced precoding techniques that can increase network spectral efficiency through point-to-point or multiuser designs.
\textcolor{black}{Continued scaling of transmit antennas will likely persist if accompanied by advances in signal processing and communication theory for multiple-input multiple-output (MIMO) systems \cite{MIMOover} and new {trustworthiness} and {sustainability} constraints.} 6G support for larger-scale MIMO installations is one of the most immediate issues. Industry today will move to extreme massive MIMO deployments, with elements numbering in the thousands.

New 6G spectrum, including the 7-16 GHz upper midband, and lower frequencies can offer substantial capacity benefits if accompanied by this larger number of antennas, new sounding techniques, and improvements in control information.
Research needs to focus on new use cases with large numbers of antennas, addressing more
complicated network-centric approaches.
Even today, the deployment of multiuser MIMO is in its infancy,
despite major theoretical advances. MIMO innovations are needed to support the increased channel state information (CSI) needs of these large arrays, especially when combined with the synchronization challenges of geographically-distributed arrays.
Uplink MIMO is another point of needed advancement.
Though uplink MIMO was included in LTE-Advanced, it has not yet received widespread adoption. This must change as the uplink is likely to become more strained as user equipment (UE)-assisted AI/ML tasks strain the network.
It is anticipated that these needed innovations will be in place before the end of the decade, making multiuser and uplink MIMO finally widely deployed in rollouts of the 6G standard.

\vspace{-0.15in}
\subsection{Intermittent Connectivity and Communication}
Mobility has always been an important topic for commercial wireless.
Management technologies have matured over the years to the current 3GPP standard's use of idle and connected states \cite{3gpp38321}.
Many of the network and transport technologies that support intermittent connectivity and communication have reached a high level of maturity as well, e.g., consider the standardization of cellular discontinuous reception techniques.

Despite the maturity, there is still ample need for research in handling intermittent connectivity and communication for 6G, particularly in {sustainability} and {digital inclusion} use cases.
One novel application in 5G has been machine-to-machine (M2M) communications, which aids in providing connected services in regions without access to broadband, e.g., smart agriculture. In 6G, M2M is expected to be more interactive and autonomous \cite{iot_commmag}. While existing network and transport layer technologies (e.g., TCP/IP, UDP, MQTT, etc.) can be leveraged to support basic needs, enhancements driving efficient use of computing, radio and energy resources (reduced packet overhead, fewer handshakes, etc.), reduced latency, and increased reliability will be required as well.

To achieve 6G sustainability goals, new protocols must be designed to enable both device and network energy savings in time, frequency carriers, and MIMO spatial domains by exploiting intermittent behavior without degrading end-to-end performance \cite{3gppnestr}.
Work is needed to support intermittent communications of trillions of intelligent ambient Internet of Things (IoT) devices, which may revolutionize multiple areas including logistics and precision agriculture. While existing short-range radio frequency identification (RFID) technology is powering tens of billions of new devices each year, 6G ambient IoT research is needed on key technologies such as ultra low-power semiconductors, high efficiency RF energy harvesters, and
dual-use waveforms and
protocols that could power sensors and enable backscatter communications.
Much of this work is underway, with full support for IoT intermittent communication expected as an early-2030s 6G feature.

\vspace{-0.15in}
\subsection{Heterogeneous Latency Requirements} \label{sec:latency}
In 5G, support for ultra-low latency applications has been enhanced through multiple approaches, e.g., decreasing the orthogonal frequency division multiplexing (OFDM) symbol period and shortening transport block time durations.
6G will further expand the vertical markets addressed and will need to support scalability to even more heterogeneous latency requirements, including applications that are more delay-tolerant but require highly reliable upper bounds on latency.

Support for applications that require ultra-low latency puts stringent processing demands on the network. 
In fact, the typical end-to-end latency is much greater than the latency on the radio interface (e.g., 10 ms vs. 1 ms), requiring improvements throughout the network.
Many of the technology developments needed for ultra-low latency are mature from a standardization point-of-view. Research is needed on the implementation of these networks for the variety of emerging 6G use cases. Much of this is rather evolutionary, as delay-tolerant IoT and high-altitude communications have prior use outside of broadband cellular networks. As a result, a wide range of 6G latency requirements will already be supported by the early 2030s.

To facilitate future demands, research will also be needed on understanding the fundamental limits of latency and achieving
them. The community's understanding of finite blocklength communication is limited to only the simplest cases (e.g., point-to-point, academic channel models). Theory is far from providing significant insight into more complicated network operation. \textcolor{black}{This is especially true when looking at multi-hop, point-to-multipoint, or multipoint-to-point deployments.}

\vspace{-0.15in}
\subsection{Millimeter Wave and Sub-Terahertz Bands} \label{sec:subTHz}
\vspace{-0.025in}
Despite the standardization of millimeter wave (mmWave) communications in 5G NR, 
widespread use of higher frequencies (e.g., above 20 GHz) is still not a reality in broadband cellular.
Research is needed to address the challenges hindering broader adoption of mmWave for scalability. Specifically, we must find physical and higher-layer techniques to broaden use cases, including overcoming the beam blockage issues.

\begin{figure}
  \centerline{\epsfig{file=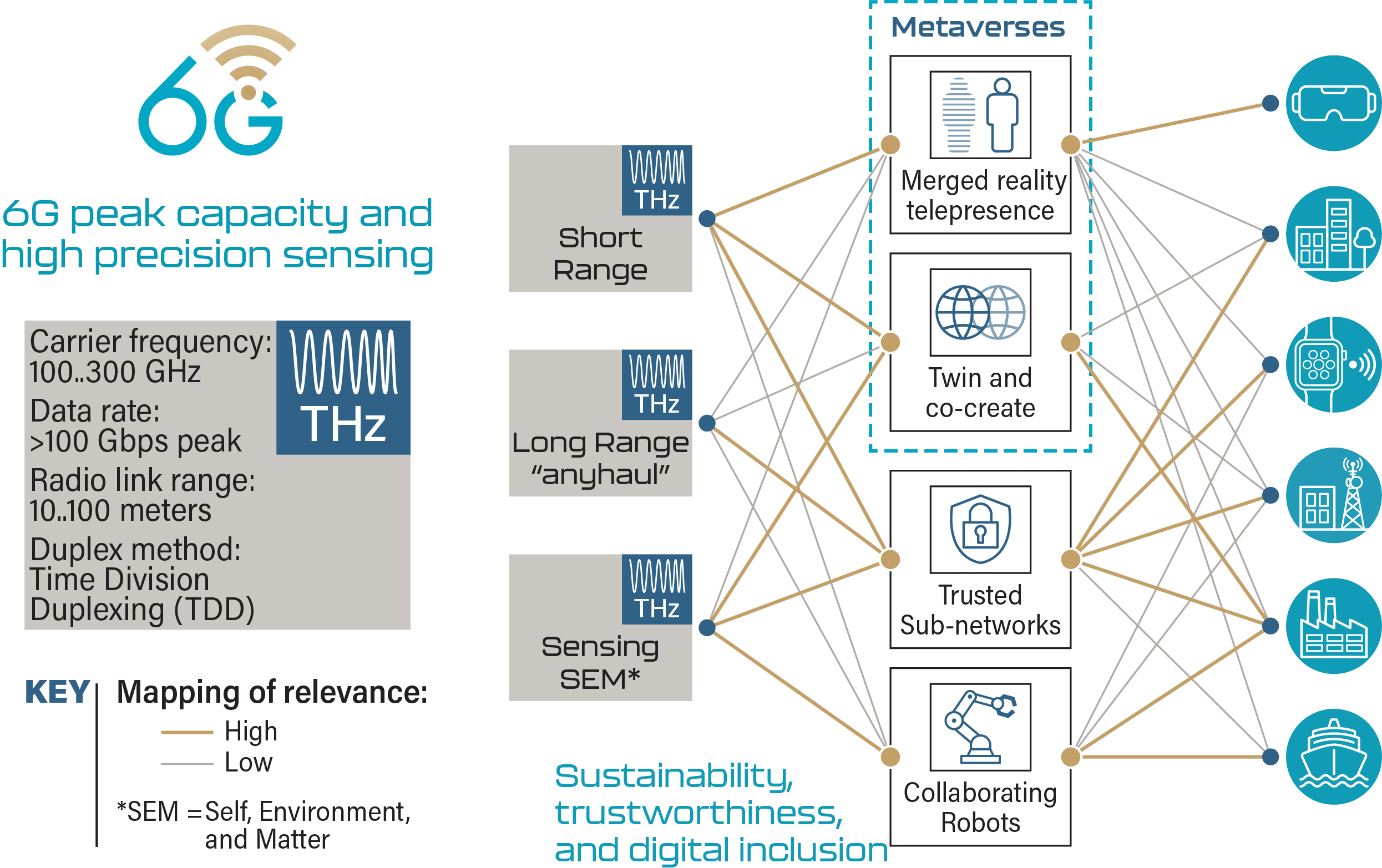,width=3.25in}} 
  \vspace{-0.1in}
  \caption{Sub-THz technology can be broadly clustered into functional enablers of access, backhaul, and sensing.
	Indicative mappings of these enablers to key use cases of 6G are shown.
	The sectoral relevance of these use cases is visualized on the right by weighted lines to key industry sectors.}
 \vspace{-0.225in}
 \label{fig:subTHz}
\end{figure}

Past mmWave research and development has shown that underutilized higher frequency bands offer large bandwidths and the ability to supply peak data rates far exceeding the best cases available in sub-6 GHz bands.
In the 6G era, sub-Terahertz (sub-THz) bands from 100-300 GHz may be used for maximum link capacity, e.g., more than 100 Gbps, and for new high-precision sensing to support sustainability and digital inclusion use cases. The use of sub-THz bands for environmental, material, and human sensing may emerge as significant application scenarios \cite{HexaX2021} (see Fig. \ref{fig:subTHz}).

Currently, development of sub-THz solutions,
with a focus on access, sensing, and backhaul, is still in the research stage, and far from practical today. Advances in midband deployments are more commercially important in the near term.
Luckily, it is likely that solutions to the existing mmWave issues will be directly portable to sub-THz. Still, due to the lack of experience in operating commercial networks on these bands, both communication theory and signal processing research are needed to obviate any trustworthiness concerns. As a result, it is expected that commercial use of these bands will be in far-term, late-2030s releases of 6G.
\vspace{-0.15in}
\section{Enhanced Network Operation}

\subsection{Non-Terrestrial Networks} \label{sec:ntn}
A major impediment to 
digital inclusion is the capital investment of deploying infrastructure in areas with limited perceived return on investment.
There are also numerous maritime and remote areas that currently have minimal viable wireless broadband solutions, which is troubling given their many untapped sustainability applications.
One potential solution is non-terrestrial networks (NTNs), operating at altitudes ranging from satellites to unmanned aerial vehicles (UAVs). NTN has received increasing momentum in 5G, but full integration is not expected until 6G.
Much recent interest has centered on low earth orbit (LEO) constellations \cite{6g_leo}, which offer lower cost of deployment, lower latency, and lower loss compared to medium (MEO) and geostationary (GEO) orbits.
Generally, coverage-per-satellite increases as altitude increases, meaning that LEO requires denser constellations. This requires managing higher system complexity and handover between satellites, leading to sustainability constraints. There is also concern that a large increase in satellites will have unforeseen effects on radio astronomy, weather forecasting, and space junk.

NTN research challenges for 6G range from physical-layer issues to higher-layer integration.
NTN operation for backbone and backhauling requires careful design and spectrum operation, particularly for scalability to urban and suburban use cases.
UE-to-NTN or UE-to-aggregator-to-NTN architectures will need cross-layer enhancements that minimize terrestrial interference and properly handle delay. It should be noted that there is a need to support \textit{all} handset types, which motivates direct handset-to-NTN access.
Beamforming transmission, interference cancellation,
and spectrum sharing/sensing with coexistence between satellites and base stations must also be studied. These solutions are expected to mature to a point of commercial deployment by the mid-2030s releases of 6G.

\vspace{-0.15in}
\subsection{Backhaul Evolution with Increased Cell Density} \label{sec:backhaul}
Increasing mobile backhaul capacity and flexibility is necessary to realize the benefits of improved performance with each network generation, enabled by innovations in the radio interface and other parts of the network.
Connecting the radio access network (RAN) with the core, the backhaul can be realized with both wireline and wireless technologies, each critical to 6G scalability.
Wireline backhaul, typically fiber-based, is capable of supporting data rates at terabits-per-second with very high reliability.
Wireless backhaul (e.g., microwave point-to-point) can use frequencies in the 6-42 GHz and 70-80 GHz ranges and will be critical to sustainability and digital inclusion objectives in reaching under-served areas.
Each new mobile network generation has sought to increase data rates, thus requiring more bandwidth and higher frequency spectrum.
Fortunately, both mmWave and sub-THz frequencies hold much potential for 6G backhaul, albeit with limited coverage.

\begin{figure}
\centerline{\epsfig{file=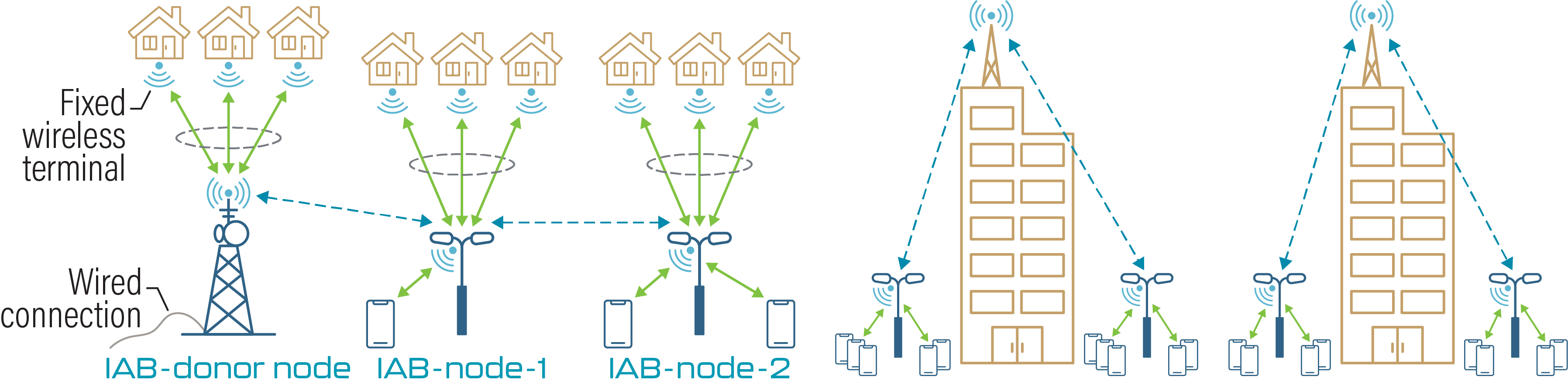,width=3.5in}}
\vspace{-0.1in}
\caption{Wireless backhaul as a coverage enhancer and service enabler (left) and as a bandwidth-efficient local capacity and bit rate booster in a macro-cell deployment with connected smaller cells (right).}
\vspace{-0.225in}
\label{fig:backhaul}
\end{figure}

Increasing user densities and higher frequencies mean that cell densities will continue to increase in 6G.
This can spike costs when using fiber backhaul, which can become sluggish in certain circumstances.
One approach to mitigate this is self-backhauling, where part of the spectrum used for access services also is used for backhaul.  Fig. \ref{fig:backhaul} demonstrates integrated access and backhaul (IAB), including the ability to provision services in areas where access has not been economical, and allowing for increased network capacities through small cells. 
Physical-layer relaying may also be used to augment backhaul.
It already has been shown that self-backhauling can be regarded as a reasonably mature technology from a specification and development perspective \cite{Ericsson2020}, and it also has been operator tested \cite{FierceWireless2020}.
However, it has not been widely deployed yet, possibly due to its higher complexity. This is an important near-term consideration given the expectation that this technology will be among early-2030s 6G rollouts.

\vspace{-0.15in}
\subsection{AI/ML-Based Network Optimization} \label{sec:AIML}
AI/ML has become an active research and technology area for 5G and 6G, aiming to optimize network operating regimes from a data-driven perspective. Recurrent neural network (RNN) architectures have demonstrated promising capabilities in signal classification, transmission coding design, traffic anomaly detection, and other tasks. Deep reinforcement learning (DRL) techniques have also shown promise for adaptive online tasks like congestion control and spectrum allocation.

5G standards development efforts have enhanced functions in the control and management planes to include AI/ML model training and inference (e.g., 3GPP Releases 17-19). 6G will evolve this to an AI-native design running on a cloud-native system infrastructure. However, according to the Next G Alliance, global realization for 6G will require concerted efforts along three dimensions \cite{NGA6Groadmap}: (1) the standard should support an open architecture to facilitate advancements; (2) operators must be convinced that AI/ML can increase their performance indicators; (3) datasets necessary for the research community for development and benchmarking must be made widely available. Given the need for several iterations of algorithm refinement on such datasets, large-scale realizations of AI-driven 6G will likely be in the mid-2030s.

There are several important considerations for algorithm development. First, the trend of making AI/ML models ``deeper'' is problematic in 6G, as minimizing protocol execution time and energy consumption are critical for scalability and sustainability. AI/ML complexity reduction can be achieved through careful control of three factors: input size, model size, and number of models. Reducing the input dimension of calls for tight integration with signal processing techniques to rapidly extract important received signal components (e.g., first/second-order I/Q sample statistics). NN sizes can be managed by architecting models around novel AI/ML paradigms like Kolmogorov-Arnold Networks (KANs), which replace fixed activation functions in traditional Multi-Layer Perceptrons (MLPs) with learnable functions, promising comparable accuracies with significantly fewer model parameters. 
The number of models can be reduced by analyzing correlations between tasks to avoid redundant training processes, e.g., an NN built for channel estimation may have an overlapping state space with one built for spectrum usage prediction.

Additionally, the impact of AI/ML on 6G robustness and trustworthiness must be carefully considered. Deep learning introduces opaqueness to end users and network operators in understanding algorithm decisions, breeding distrust. 6G must integrate explainable AI (XAI) techniques alongside models to promote transparency. Approaches being considered by industry include feature analytics (i.e., determining which inputs are driving an NN's decision) and contrastive explanations (i.e., revealing the difference in reward between a DRL's chosen action vs. other possibilities). Moreover, AI/ML will open up susceptibility to more cyberattacks, including data and model poisoning, inversion, and evasion attacks attempting to compromise 6G performance. One line of defensive strategies being considered is having independent algorithms devoted to intrusion detection, with 6G operations bypassing AI/ML decisions when the network is in a compromised state. Another set of strategies involve enhancing the resilience of the AI/ML algorithms themselves, e.g., through adversarial training.
\vspace{-0.15in}
\section{Computing and Communications Convergence} \label{sec:conv}

\subsection{Open Interface} \label{sec:openinterface}
Open interfaces, with contributions from organizations like 3GPP and O-RAN Alliance, have enabled a diverse group of suppliers for devices and networks.
They also provide a standardized mechanism for different components of the mobile network to interoperate. This creates the possibility of specialized and optimized design for individual components and potential sourcing from different vendors. The RAN and core have been connected with an open interface since 3G. The trend towards open interfaces has continued with the separation of the RAN into specialized component functions in 5G. This is expected to play an even more important role in 6G technology, with a renewed focus on trustworthiness.

\begin{figure}
\centerline{\epsfig{file=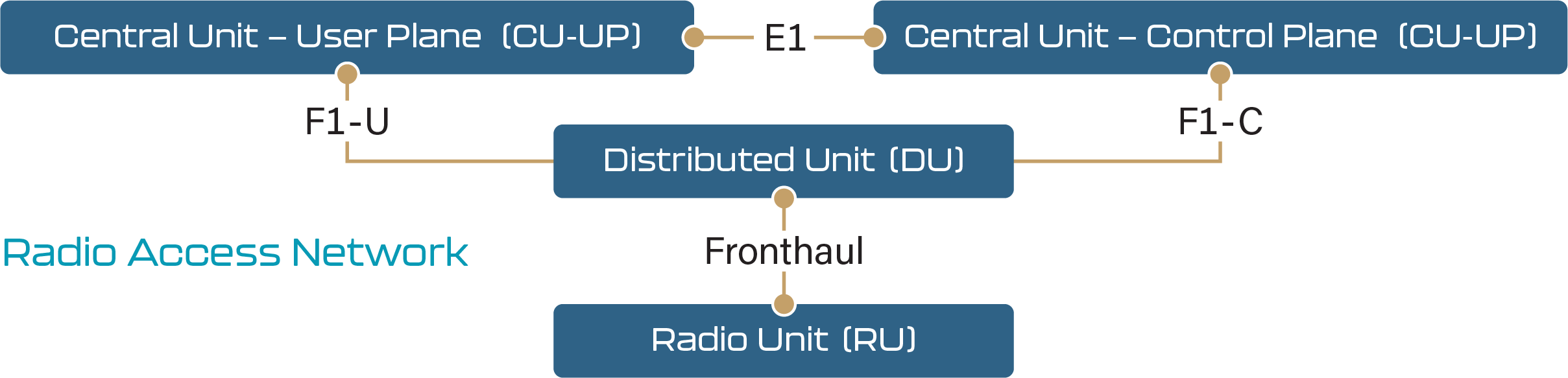,width=3.25in}}
\vspace{-0.1in}
\caption{Open interfaces in 5G RAN which must be reconsidered for 6G.}
\label{fig:openinterface}
\vspace{-0.2in}
\end{figure}

Fig. \ref{fig:openinterface} gives an illustration of 5G RAN with recently introduced open interfaces.
It contains (i) a vertical split of protocol stack functions across a central unit (CU), distributed unit (DU), and radio unit (RU) of a RAN, which can be located in different parts of the deployment region, as well as (ii) a horizontal separation of control and user planes. The CU utilizes general-purpose computing, the DU relies on signal processing and accelerators, and the RU hosts power amplifiers and radio antennas.
With open interfaces between these functions, providers specializing in specific components, e.g., power amplifiers for the RU or signal processing for the DU, can work to develop best-in-class solutions.

The architecture of open interfaces for 6G is an important question, as it must support new use-cases including cyber-physical systems, immersive experiences, and NTN integration. The availability of testbeds to prove out solutions based on open interfaces will be critical. Examples of organizations providing such testbeds include the Telecom Infrastructure Project, the SONIC Lab in the UK, and projects such as Colosseum in the US. Additionally, 6G will likely promote a service-based architecture (SBA) as an alternative to traditional fixed function, proprietary hardware, potentially revolutionizing digital inclusion in developing economies by opening opportunities for deployment through open interfaces \cite{ziegler20206g}. As a result, completing the transition from 5G open interfaces to the full 6G vision is likely a farther term, later-2030s feature.

Open interfaces are needed to build application programming interfaces (APIs) for developers, to provide them better access to 6G capabilities. Examples include APIs for sensing and positioning, energy-efficient workload balancing, monitoring of connectivity in underserved areas, and creating guaranteed slices for public safety and public services. An important design problem is moving from a network-centric view, where focus is on interfaces within the network, to an application-centric view, where focus is on exposing the value created by the network to application developers. This design approach will be an important building block to realizing economic, scalability and digital inclusion benefits of 6G.

\vspace{-0.15in}
\subsection{Platforms for Wireless Service}
Advances in computing infrastructure have enabled solutions previously developed on purpose-built hardware to move to software running on general-purpose platforms. This trend has accelerated with the adoption of cloud technologies for application workloads, and has resulted in the ability to develop services with higher scalability.
The proliferation of cloud platforms has provided an additional impetus for open interfaces.
However, the deployment of RANs on cloud platforms also has highlighted the need for custom hardware accelerators for some critical functions with strict latency and high reliability requirements.
Examples include channel decoders, equalizers, and processing for MIMO systems.

These technological developments have resulted in increased programmability of the mobile network, allowing service providers to cater to a wider range of requirements.
This has enabled 5G mobile networks to serve heterogeneous use cases for vertical industries, enterprises, and governments in addition to traditional voice and mobile broadband.
The applications that need to be addressed will further expand to new verticals with 6G, accentuating the importance of innovations in network configurability that will allow them to adapt to different traffic patterns and deployment scenarios while addressing the focus areas of scalability and sustainability.

The evolution to more programmable networks yields opportunities for AI/ML to enable more intelligent and adaptable 6G networks as well.
Work is also needed on combining computing-driven implementations with advanced APIs, which enable developers to create applications and services that depend on wireless connectivity.
A cognitive global network platform accessible through APIs will hide the underlying complexity while enabling the provision of advanced 6G applications and services that require connectivity, and may in addition integrate computing, sensing, and control capabilities. This fully integrated global wireless service platform is expected to be a far-term 6G feature, in the late-2030s.

\vspace{-0.15in}
\subsection{Edge/Fog Computing} \label{sec:edgefog}
Cloud computing faces fundamental scalability challenges in managing the exponential growth in edge devices and data-hungry applications.
This motivated edge computing as an alternative, leveraging increasingly powerful processing capabilities of edge devices to handle computation services closer to the point of origin.
These devices often are heterogeneous in their on-board resources, which becomes a key consideration for optimizing task workloads and sustainability objectives.

\begin{figure}
\centerline{\epsfig{file=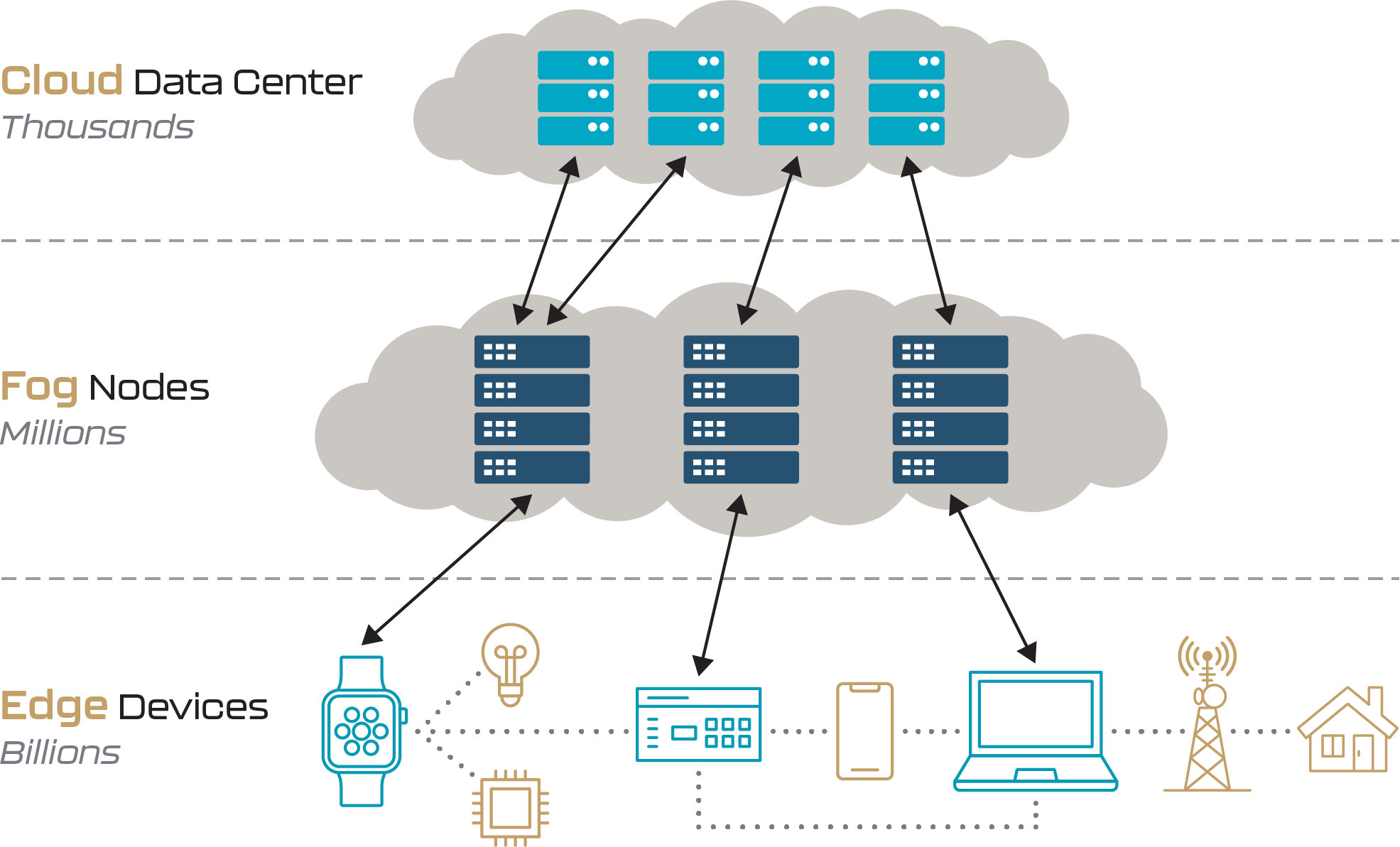,width=3in}}
\vspace{-0.15in}
\caption{Fog computing aims to orchestrate computing resources along the cloud-to-things continuum, providing scalability for 6G services.}
\vspace{-0.225in}
\label{fig:fog}
\end{figure}

Edge and cloud computing can be viewed as the ``extremes'' of a processing/storage model.
While the cloud provides an ideal configuration for complex data processing tasks that can tolerate some delay, the edge can service more localized tasks that have stringent delay and privacy requirements.
Aiming for a hybrid, fog computing considers intelligently orchestrating compute, storage, and networking services between edge/IoT devices, in-network computing (e.g., routers, edge servers), and the cloud (see Fig. \ref{fig:fog}). With fog, the 6G mobile system will undergo a fundamental transformation into a wide-area cloud, inherently delivering pervasive computing services \cite{li20226g}.

Fog computing must manage many tasks with diverse service requirements while satisfying scalability, sustainability, and trustworthiness objectives.
The proliferation of data-intensive AI/ML tasks -- both for user applications and for network control (Sec.~\ref{sec:AIML}) -- is accelerating this urgency for 6G. In this respect, federated learning (FL) has emerged as a popular paradigm for distributed AI/ML, where edge devices iteratively train local models which are periodically aggregated by a cloud server. An advantage of FL is that it obviates raw data sharing over the network, providing trustworthiness advantages. However, the client-server star topology of FL encounters significant challenges as the network systems and models scale up in size and heterogeneity.

To overcome this, research in ``fog learning'' is aiming to intelligently distribute AI/ML data processing and model aggregations across the network \cite{Wang-Hosseinalipour-Aggarwal2023}. A key principle of fog learning is to leverage the proliferation of direct device-to-device (D2D) and M2M communications for dynamic resource pooling and local node collaboration. Preliminary results have shown how intelligent edge-fog-cloud coordination can lead to substantial improvements in service quality, latency, and energy consumption metrics for AI/ML services. On the other hand, local data and model sharing in fog learning disrupts privacy safeguards provided by FL. Additional approaches are therefore needed to preserve trustworthiness, e.g., through privacy-preserving data and model encodings, or restricting cooperation to user-selected networks of trust. These techniques are expected to mature for a mid-2030s feature of 6G, aligned with the AI/ML services they will enable.
\vspace{-0.1in}
\section{Rethinking Network Deployment}

\subsection{Spectrum Sharing} \label{sec:spectrumsharing}
Historically, spectrum sharing has employed static, semi-dynamic, or autonomous best-effort mechanisms. An operator migrating its carrier from 4G to 5G is an example of static sharing. A semi-dynamic sharing example is the US Citizen’s Broadband Radio Service (CBRS) sharing between federal incumbent users, priority access licensees (PAL), and general authorized access (GAA). Wi-Fi’s Listen-Before-Talk (LBT) is an example of autonomous best-effort spectrum sharing.

More recently, 3GPP introduced a dynamic spectrum sharing (DSS) feature for 5G NR that enables the parallel use of 4G LTE and 5G NR in the same band.
DSS as defined by 3GPP should be further developed for 6G to cover more/all possible dimensions, including across locations, frequencies, time, users, and applications \cite{3gpprel17work}, and requiring new levels of trustworthiness. Potential designs include advanced spectrum sensing, interference analysis and avoidance, AI/ML-based sharing control, inter-system/inter-user direct signaling for spectrum sharing, and hierarchical spectrum management. Such techniques for enhanced DSS are expected to appear in the initial releases of 6G in the early 2030s.

Nevertheless, there are many design challenges and practical considerations for DSS technology in 6G. For example, in certain radio environments, the coordination between two systems may not be possible or may be overly complex. Moreover, legacy devices and networks exist that cannot be upgraded to support DSS, especially in many IoT markets. One promising line of work is joint spectrum sensing and resource allocation, aiming to maximize throughput of secondary users employing DSS protocols while minimizing interference with uncontrollable primary users. More generally, a holistic effort among regulators, infrastructure operators, device users, and the academic community is necessary.

\vspace{-0.15in}
\subsection{Shared Infrastructure} \label{sec:sharedinfra}
\begin{figure}  \centerline{\epsfig{file=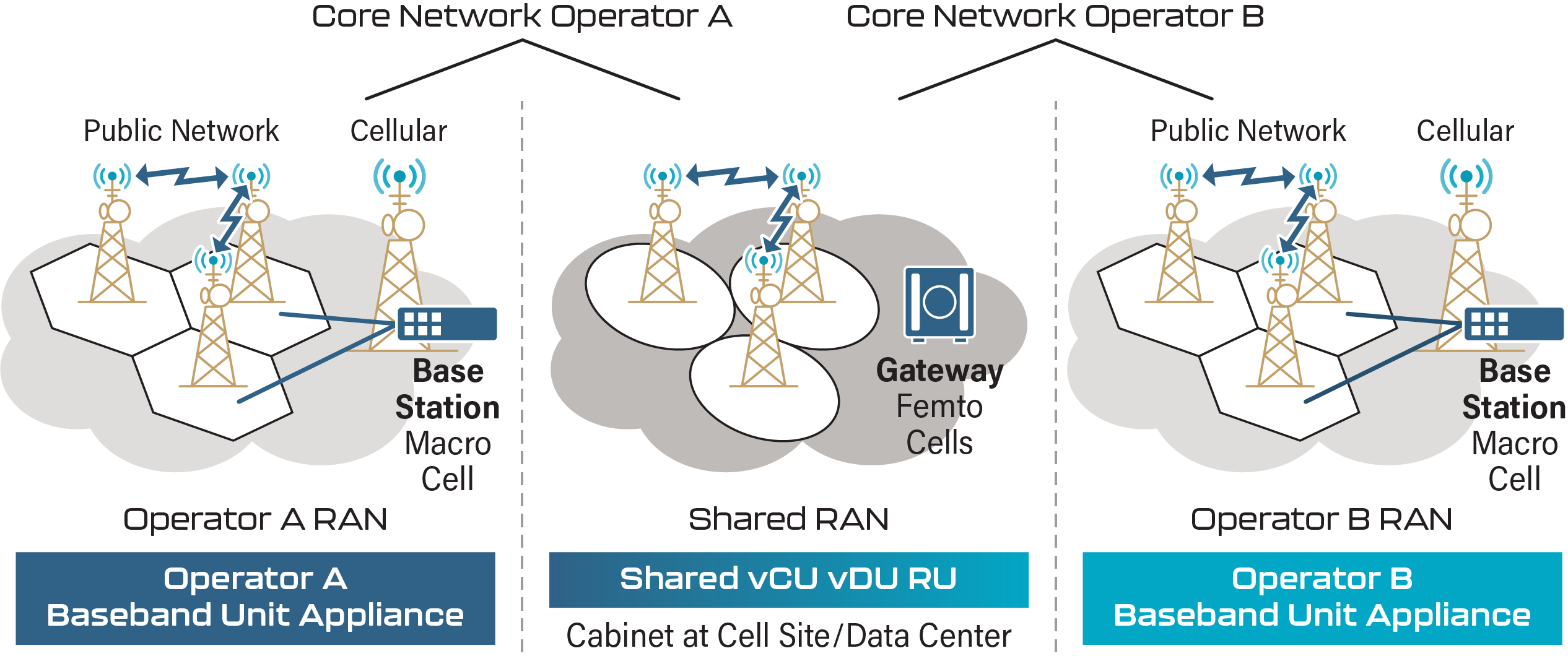,width=3.25in}}
  \vspace{-0.1in}
\caption{Operators who utilize shared network infrastructure can execute network functions as highly disaggregated and distributed microservices.} \label{fig:sharedinfra}
  \vspace{-0.225in}
\end{figure}

End-to-end virtualization is critical to supporting multi-tenant, multi-operator shared environments.
5G's disaggregated RAN (Sec. \ref{sec:openinterface}) is a good example of today's evolution towards virtualized, shared environments.
Cloud-based orchestration and management principles can be applied to fully virtualized SBA-based RANs envisioned for 6G.
With versatile cloud platforms that enable developers to build, deploy, and execute server-side applications swiftly, operators using shared network infrastructure can run network functions as highly disaggregated and distributed microservices (see Fig. \ref{fig:sharedinfra}).

Though the transition to virtualized RAN is well under way, there are still numerous research challenges, particularly in the presence of the evolving focus area constraints.
For example, the interdependence of network functions remains a barrier to achieving fully cloud-native implementations. The convergence of computing, communication, and control (Sec. \ref{sec:conv}) also presents issues for infrastructure sharing. As a result, shared infrastructure with virtualized layers is likely a far-term, late-2030s feature of 6G. One key area of exploration for 6G is how telemetry data that can be gathered by operators on shared infrastructure with virtualized layers can be used to improve sustainability. Non-real time and near-real time RICs enable the combination of network platform and application telemetry. Research should consider how these RICs can be used to increase spectral efficiency and usage of deployed radio resources with active antenna management and orchestration, improving scalability by dynamically allocating network function resources as network conditions change.

\vspace{-0.15in}
\subsection{Enhanced Internetworking and Security}
The 6G scalability focus area requires better, more seamless coverage and handover across different environments.
These environments include different indoor and outdoor settings serviced by multiple protocols, vendors, and systems.
To this end, standards specifications, technical architectures and operational frameworks are needed to facilitate a high degree of internetworking and user experience compatibility.

Of particular importance is the interoperability of WiFi and 6G. WiFi 6 was already designed to be more transparent to 5G than in previous generations, providing a starting point for full internetworking in 6G by the mid-2030s. To accomplish this, research must explore establishing a common identity for a user, device, or other IoT node that allows consistent bandwidth and latency policies, security enforcement and analytics, networking roaming privileges, and other attributes applied to that user/device. Regardless of the access method (WiFi or 6G), the resultant communications must then adhere to the required security functions, and the resultant security analytics and alerts should be processed consistently. Further, future networks should support open roaming between technologies without requiring user input. In particular, devices with access to multiple radio networks should automatically optimize selection based on multiple factors, e.g., power profiles, bandwidth, latency, and reliability metrics. They should also employ techniques that bond together information flowing over the multiple networks for higher-layer multipath connectivity.

Another key consideration is that 6G is likely to be deployed in a post-quantum world. Studies are needed on quantum interoperability, with a rethinking of the attack surface. We must not deploy technological advancements that improve spectral efficiency but inadvertently introduce new vulnerabilities.
\vspace{-0.15in}
\section{Regulatory and Public Policy Challenges}
\label{sec:patnership}\label{sec:standards}
As mentioned in Sec.~\ref{sect_intro}, 6G pre-standardization activities have initiated in the last few years, with various regional groups around the world bringing together stakeholders.
Building upon past success, the International Telecommunications Union (ITU)-R (Radiocommunication Sector) is currently working towards the IMT-2030 requirements and evaluation, with the specification to be completed in the 2030 timeframe. 3GPP work on 6G commences in Release 20 with a study item, and the first 6G specification might come in Release 21.

There are increasing stakeholders in the communications community responsible for device manufacturing, network operation, computing platforms, and other aspects.
Cooperation will be needed to meet the IMT-2030 goals and generate consensus on spectrum bands. 6G bands of interest can roughly be grouped into sub-1GHz, mid-band, upper mid-band, mmWave, and sub-THz, including all spectrum ranges from 5G as well as new bands for exclusive 6G use.
The success of mid-band wide area deployments in 5G should be further be enhanced in 6G by having policymakers consider the upper mid-band spectrum of roughly 7-16 GHz for new deployments.

Additionally, to maximize socio-economic benefit, a major effort is needed to identify spectrum bands that can be shared between federal and commercial users (e.g., CBRS in Sec. \ref{sec:spectrumsharing}). Rules that account for practical impacts should be defined for spectrum sharing between different commercial technologies. At the same time, legacy non-wireless broadband systems often are using equipment that is decades old. This concern was brought to the media’s attention during the radar altimeter debate between 5G operators. Legacy users must be encouraged to update their equipment, and new signal processing and waveform-based solutions must protect such users. Often, the constraints will be \textit{geographic}, with specific locations requiring restrictions. Spatial processing using multiple antennas and/or new array structures could both alleviate the worries of legacy users and increase network throughput.

As discussed, many 6G innovations are expected to arise from the proliferation of open interfaces between hardware and software (see Sec. \ref{sec:openinterface}).
Over the last few years, the O-RAN Alliance has been bringing together a large community around standardization, open software development, and implementation testing/integration of RAN technologies, with an eye toward virtualization and interoperability (see Sec. \ref{sec:sharedinfra}). Collaboration within and among such organizations, as well as alignment with government strategies, is essential.

Finally, government partnerships will likely be necessary for the world's digital inclusion issues. Terrestrial and NTN deployments aimed at these challenges will require technological advancement, policy solutions, and global consensus. Consensus on supply chain improvements will also be critical for trustworthiness: 5G manufacturing issues related to COVID-19 necessitate creation of a resilient hardware pipeline for 6G.
\vspace{-0.25in}
\section{Concluding Remarks}
We discussed 12 technology innovations that will coalesce into 6G standards, ranging from advancements in computing to radio access.
Focused research in the identified technologies is poised to make significant advancements in scalability, sustainability, trustworthiness, and digital inclusion.

\vspace{-0.15in}
\bibliographystyle{IEEEtran}
\bibliography{bibs/6Gtaxonomy}

\vspace{-0.1in}
\vspace{-0.4in}
\begin{IEEEbiographynophoto}{Christopher G. Brinton}
(cgb@purdue.edu)
is Elmore Associate Professor of ECE at Purdue University.
\end{IEEEbiographynophoto}
\vspace{-0.5in}
\begin{IEEEbiographynophoto}{Mung Chiang}
  (chiang@purdue.edu)
is President of Purdue University.
\end{IEEEbiographynophoto}
\vspace{-0.5in}
\begin{IEEEbiographynophoto}{Kwang Taik Kim}
  (kimkt@purdue.edu)
is Research Assistant Professor of ECE at Purdue University
\end{IEEEbiographynophoto}
\vspace{-0.5in}
\begin{IEEEbiographynophoto}{David J. Love}
  (djlove@purdue.edu)
is Nick Trbovich Professor of ECE at Purdue University.
\end{IEEEbiographynophoto}
\vspace{-0.5in}
\begin{IEEEbiographynophoto}{Michael Beesley}
  (mbeesley@cisco.com)
is Vice President/Chief Technology Officer at Cisco Systems.
\end{IEEEbiographynophoto}
\vspace{-0.5in}
\begin{IEEEbiographynophoto}{Morris Repeta}
  (morris\_repeta@dell.com)
is Director of Advanced Wireless Technology at Dell Technology.
\end{IEEEbiographynophoto}
\vspace{-0.5in}
\begin{IEEEbiographynophoto}{John Roese}
  (john.roese@dell.com)
is President and Global Chief Technology Officer at Dell Technology.
\end{IEEEbiographynophoto}
\vspace{-0.5in}
\begin{IEEEbiographynophoto}{Per Beming}
  (per.beming@ericsson.com)
is Head of Standard and Industry Initiatives at Ericsson.
\end{IEEEbiographynophoto}
\vspace{-0.5in}
\begin{IEEEbiographynophoto}{Erik Ekudden}
  (erik.ekudden@ericsson.com)
is Senior Vice President and Chief Technology Officer at Ericsson.
\end{IEEEbiographynophoto}
\vspace{-0.5in}
\begin{IEEEbiographynophoto}{Clara Li}
  (clara.q.li@intel.com)
is Senior Principal Engineer at Intel.
\end{IEEEbiographynophoto}
\vspace{-0.5in}
\begin{IEEEbiographynophoto}{Geng Wu}
  (geng.wu@intel.com)
is Intel Fellow at Intel.
\end{IEEEbiographynophoto}
\vspace{-0.5in}
\begin{IEEEbiographynophoto}{Nishant Batra}
  (nishant.batra@nokia.com)
is Chief Strategy and Technology Officer at Nokia.
\end{IEEEbiographynophoto}
\vspace{-0.5in}
\begin{IEEEbiographynophoto}{\textcolor{black}{Amitava Ghosh}}
  \textcolor{black}{(amitava.ghosh@nokia.com)
is Nokia Fellow and Head of Radio Interface Group at Nokia.}
\end{IEEEbiographynophoto}
\vspace{-0.5in}
\begin{IEEEbiographynophoto}{Volker Ziegler}
  (volker.ziegler@nokia.com)
is Senior Advisor and Chief Architect at Nokia.
\end{IEEEbiographynophoto}
\vspace{-0.5in}
\begin{IEEEbiographynophoto}{Tingfang Ji}
  (tji@qti.qualcomm.com)
is Vice President at Qualcomm.
\end{IEEEbiographynophoto}
\vspace{-0.5in}
\begin{IEEEbiographynophoto}{Rajat Prakash}
  (rprakash@qti.qualcomm.com)
is Senior Director at Qualcomm.
\end{IEEEbiographynophoto}
\vspace{-0.5in}
\begin{IEEEbiographynophoto}{John Smee}
  (jsmee@qti.qualcomm.com)
is Senior Vice President at Qualcomm.
\end{IEEEbiographynophoto}

\end{document}